\documentclass[%
 reprint,
  superscriptaddress,
 amsmath,amssymb,amsfonts,
 aps,
 prl,
 floatfix,
 longbibliography
]{revtex4-2}

\usepackage{bm}

\usepackage[pdftex]{graphicx} \graphicspath{{}}
\usepackage{float} \usepackage{color}

\usepackage[pdftex,colorlinks=true]{hyperref}
\hypersetup{
  colorlinks=true,
  linkcolor=blue,
  urlcolor=cyan,
}

\usepackage{mathtools}

\DeclarePairedDelimiter\ket{\lvert}{\rangle}
\DeclarePairedDelimiterX\braket[2]{\langle}{\rangle}{#1 \delimsize\vert #2}
\DeclarePairedDelimiterX\expval[3]{\langle}{\rangle}{#1 \delimsize\vert #2  \delimsize\vert #3}
\DeclarePairedDelimiterX\sexpval[1]{\langle}{\rangle}{#1}

\usepackage{siunitx}

\newcommand{\vect}[1]{\mathbf{#1}}

\begin{document}

\title{Disentangling Pauli blocking of atomic decay from cooperative radiation and atomic motion in a 2D Fermi gas}

\author{Thomas Bilitewski}
\affiliation{JILA, National Institute of Standards and Technology and Department of Physics, University of Colorado, Boulder, CO, 80309, USA}
\affiliation{Center for Theory of Quantum Matter, University of Colorado, Boulder, CO, 80309, USA}
\author{Asier Pi\~neiro Orioli}
\affiliation{JILA, National Institute of Standards and Technology and Department of Physics, University of Colorado, Boulder, CO, 80309, USA}
\affiliation{Center for Theory of Quantum Matter, University of Colorado, Boulder, CO, 80309, USA}
\author{Christian Sanner}
\author{Lindsay Sonderhouse}
\author{Ross B. Hutson}
\author{Lingfeng Yan}
\author{William R. Milner}
\author{Jun Ye}
\affiliation{JILA, National Institute of Standards and Technology and Department of Physics, University of Colorado, Boulder, CO, 80309, USA}
\author{Ana Maria Rey}
\affiliation{JILA, National Institute of Standards and Technology and Department of Physics, University of Colorado, Boulder, CO, 80309, USA}
\affiliation{Center for Theory of Quantum Matter, University of Colorado, Boulder, CO, 80309, USA}

\date{\today}

\begin{abstract}
 The observation of  Pauli blocking of atomic spontaneous decay via direct measurements of the atomic population requires the use of long-lived atomic gases where quantum statistics,  atom recoil  and cooperative  radiative processes are all relevant. We develop a theoretical framework capable of simultaneously accounting for all these effects in a regime where prior theoretical approaches based on semi-classical non-interacting or interacting frozen atom approximations fail.  We apply it to atoms in a single 2D pancake or arrays of pancakes featuring an effective  $\Lambda$ level  structure (one excited and two degenerate ground states).  We identify a parameter window in which a factor of two extension in the atomic lifetime clearly attributable to Pauli blocking  should be experimentally observable in deeply degenerate gases with  $\sim  10^{3} $ atoms. Our predictions are supported by observation of a number-dependent excited state decay rate on the ${}^{1}\rm{S_0}-{}^{3}\rm{P_1}$  transition in $^{87}$Sr atoms.
\end{abstract}

\maketitle

{\it Introduction.---}Spontaneous emission  emerges  from the interaction of the dipole moment of an atom  with the vacuum of electromagnetic field modes. It depends on the atomic internal structure but also  on  the density of final states of the joint atom-photon system, which can be modified  by external means  as widely demonstrated in cavity QED~\cite{berman1994cavity,RevModPhys.73.565,Kimble_1998,Haroche_1989,Haroche_1999,Purcell_1946,Kleppner_1981,Goy_1983,Hulet_1985,Mataloni_1987,Haroche_1987,Thompson_1992,Hood1447,Rempe_1999,Miller_2005,Walther_2006,Fink2008,Mirhosseini2019,Haroche2020} and waveguide QED~\cite{RevModPhys.89.021001,Fan_2005,Tey2008,Schuller2010,Delsing_2011,Hoi2015,Maser2016,Liu2017,Mirhosseini2018,RevModPhys.93.025005,Burkard2020,Blais_2004,Wallraff2004,Clerk2020,GU20171,Yablonovitch_1987,Yablonovitch_1988,Yablonovitch_1989,Igeta_1991} systems. 

The density of final states can also be modified by Fermi statistics and Pauli blocking  of the available external motional states  into which an electronically excited  atom  can decay  \cite{HELMERSON_90,PhysRevA.52.3033,PhysRevA.58.R4267,PhysRevLett.82.4741,Jin_1998,Ketterle_2001,Busch_1998,Shuve_2009,Busch_2009,Zoller_2011}. 
Observation of  Pauli blocking of radiation has  been difficult due to the complex interplay of cooperative effects and  atomic motion. Cooperative effects emerge  from  the virtual exchange of photons with other atoms, which results in dipolar interactions that can enhance or suppress the radiative  decay rate of the system~\cite{PhysRevA.94.023612,PhysRevA.2.883,PhysRevA.47.1336,PhysRevA.55.513,PhysRevA.81.053821,JoMO_Kaiser_2011,PhysRevLett.108.123602,PhysRevLett.116.083601,PhysRevA.93.023407,Bromley2016,PhysRevLett.117.073003,Weiss_2018,Kaiser_2017},  especially  at the high densities required to observe Pauli blocking. At the same time the emission process directly couples the motional to the internal degrees of freedom as energy and momentum is exchanged between the atoms and photons. Thus, there is a competition between the recoil momentum and the extent of the Fermi sea that blocks spontaneous emission.

Recent experiments have for the first time  observed Pauli blocking through  measurements  of light scattered by atomic  ensembles ~\cite{sanner2021pauli,margalit2021pauli,deb2021observation}. There, the mentioned  undesirable competing effects  were minimized by performing  angular resolved measurements to select low momentum transfer  processes and by using a far detuned probe to minimize cooperative dipolar processes and suppress the number of excited atoms. 

However, an observation of enhanced life times due to Pauli blocking by direct measurements of the excited state population has yet to be demonstrated. Resonantly exciting a significant fraction of atoms to the excited state to subsequently measure their decay may result in significant dipolar interaction effects in striking contrast to scattering experiments. Moreover, working on a slow transition enabling time-resolved observation of the excited state population poses the challenge that radiative decay rates and cooperative effects become comparable with the Fermi energy and associated motional degrees of freedom which all need to be taken into account. 

\begin{figure}
\includegraphics[width=\columnwidth]{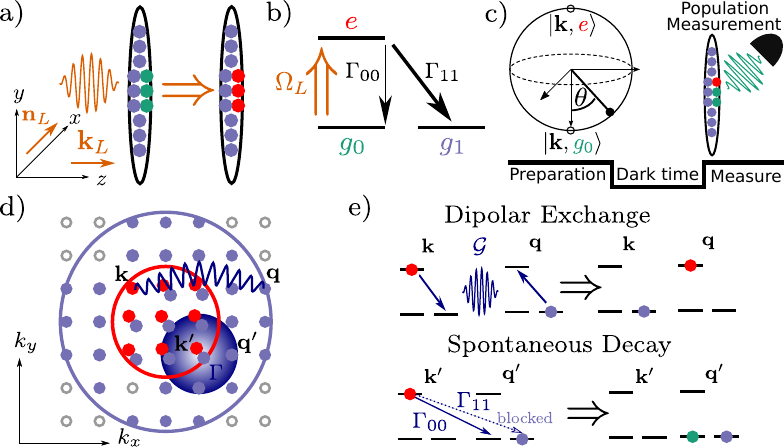}
\caption{a) A 2D cloud of atoms is optically excited by a laser pulse  with  Rabi frequency  $\Omega_L$  and  wave-vector $\mathbf{k}_L$ propagating perpendicularly to the  2D plane and linearly polarized ($\mathbf{n}_L$) along $x$.
b) Internal level structure  ($\Lambda$-system), with one excited state $e$ and two ground states $g_{\alpha}$. The single-particle decay rate for $e\rightarrow g_\alpha$ is $\Gamma_{\alpha\alpha}$. c) Protocol: after state preparation, atoms evolve freely and decay during a dark time $t$, followed by population measurement. d) In-plane momentum space $(k_x,k_y)$ with two Fermi seas (red/blue circles). Filled circles denote occupied states. Interaction processes involving virtual exchange of photons between atoms at $\vect{k}$ and $\vect{q}$ are depicted as a wiggly line, single particle spontaneous decay proportional to $\Gamma$ are illustrated as the circular region (blue shading) around $\vect{k}^{\prime}$.  e) Processes included in the master equation. Top: Dipolar exchange between atoms in momenta $\vect{k}$ and $\vect{q}$. Bottom: spontaneous decay from $\vect{k}^{\prime}$ to $\vect{q}^{\prime}$.
\label{fig:Fig1}}
\end{figure}

In this work, we develop a theoretical framework based on a  master equation (ME) formulated in momentum space capable of describing the full dipolar dynamics of optically excited atoms confined in two dimensions (see Fig.~\ref{fig:Fig1}), or in stacks of two-dimensional pancakes. 
Our framework significantly advances the applicability of theory into the interacting quantum degenerate regime, where prior approaches fail since they either account for Pauli blocking in a semi-classical non-interacting setting \cite{Busch_1998,Busch_2009,Shuve_2009}, or include interactions, but cannot account for atomic  recoil or  Pauli blocking in a natural way (frozen atom coupled dipole models) \cite{PhysRevA.94.023612,PhysRevA.2.883,PhysRevA.47.1336,PhysRevA.55.513,PhysRevA.81.053821,JoMO_Kaiser_2011,PhysRevLett.108.123602,PhysRevLett.116.083601,PhysRevA.93.023407,Bromley2016,PhysRevLett.116.083601,Kaiser_2017,Weiss_2018,PhysRevLett.117.073003}.

Our key finding is that a highly imbalanced ultra-cold Fermi gas   excited by a  resonant $\pi$ pulse  (which suppresses coherences) can feature  at $T/T_F\sim 0.1$ (with $T_F$ the Fermi temperature)  up to 50$\%$ Pauli suppression at peak densities of $10^{14}\, \mathrm{cm}^{-3}$ in  a parameter regime  where cooperative effects only affect the lifetime weakly. 
Our predictions are consistent with measurements on the ${}^1S_0-{}^3P_1$ transition in fermionic  ${}^{87} \mathrm{Sr}$  at  $T/T_F=0.6$ with a natural  lifetime  of  $\Gamma^{-1}= 21.3 \, \mu\mathrm{s}$. The experimental results not only demonstrate  the relevance  of our theory model, but also  help validate its capability to capture the essential physics  in a  complex many-body regime where exact numerical calculations are intractable. It also stimulates future experimental efforts to observe Pauli blocking through  direct lifetime measurements, which could have important implications in atomic clocks.

{\it Model.---}%
We analyze  first the case of a  Fermi gas in a single pancake in the regime where only the ground state harmonic oscillator mode $n_{0,z}$ is occupied, but motion is allowed in $x$ and $y$ (Fig.~\ref{fig:Fig1}a).
For simplicity, we  work with two-dimensional plane-waves in $x,y$ as our single-particle atomic basis, labelled by the momentum $\mathbf{k}=(k_x,k_y)$.

We consider  an effective $\Lambda$ type internal level structure (Fig.~\ref{fig:Fig1}b) with two internal ground states as the minimal system allowing strong Pauli blocking even under full excitation of one of the states, but note that our results can be straightforwardly extended to more general multilevel systems. For specificity  we focus on the ${}^1S_0\,(F=9/2)$ to ${}^3P_1\,(F=11/2)$ transition of ${}^{87}\text{Sr}$, with the $m_F=-9/2$ excited state as $e$ and the $m_F=-9/2,-7/2$ ground states as $g_0$ and $g_1$, respectively, and set the  quantization axis along $x$.  
We assume the presence of a magnetic field large enough to suppress transitions to other levels (which are omitted in the figure). The atoms are initially in an  incoherent mixture with  $N_0$ atoms in $g_0$ and $N_1$ in $g_1$. This configuration features  $\pi$  and $\sigma^-$ polarized  decay, at rates $\Gamma_{00}=2/11 \Gamma$ and  $\Gamma_{11}=9/11 \Gamma$, respectively, with $\Gamma^{-1}= 21.3 \, \mu\mathrm{s}$ \cite{Nicholson2015}.

Atoms are excited by a short laser pulse with pulse area $\theta$ and then let to evolve and decay for some time $t$ in the dark after which the total excited state population is measured (Fig.~\ref{fig:Fig1}c). We focus on the effective decay rate $\gamma_{\mathrm{eff}}(\theta) = \lim_{t\rightarrow0} \dot{N}_{ee}(t)/N_{ee}(0)$ obtained at initial time for the total number of excitations $N_{ee}(t)$. While the decay rate can change with time, in the cases discussed here the decay at early times is  well approximated by an exponential decay with rate $\gamma_{\mathrm{eff}}$.

 The  initial excitation laser with wave number $\mathbf{k}_L$ is  propagating  along the strongly confined $z$ direction and linearly polarized ($\mathbf{n}_L$) along $x$ so that it   only excites  $g_0$ atoms  to $e$ (Fig.~\ref{fig:Fig1}b).  Due to the strong confinement along $z$, the motional state does not change during excitation, so the  Rabi pulse transfers a  $g_0$ atom with momentum  $\vect{k}$  into the superposition $\cos(\theta/2) \ket{g_0,\mathbf{k},n_{0,z}} + \sin(\theta/2) \ket{e,\mathbf{k},n_{0,z}}$ (Fig.~\ref{fig:Fig1}c), where $\theta = \Omega_L t$ with the Rabi coupling $\Omega_L$ of the $g_0$-$e$ transition. The pulse is assumed to be  fast so we can  ignore any interactions during it.

To describe the  dynamics  during the dark time we start from the usual atom-light Hamiltonian and perform a Born-Markov approximation leading to a multilevel master equation (ME)  with dipolar interactions~\cite{PhysRevA.101.043816}.   We further  assume that momentum-changing interactions are negligible.
The latter approximation is justified at   early times by the initial condition, which does not contain coherences between states of different momenta $\mathbf{k}'\neq \mathbf{k}$.
For the atomic density matrix this leads \cite{supplemental} to the following ME, $\dot{\hat{\rho}} = -i \left[\hat{H},\hat{\rho} \right] + \mathcal{L}(\hat{\rho})$ with
\begin{align}
    \hat{H} &=  \sum_{\alpha, \beta} \bigg(\sum_{\mathbf{k},\mathbf{q}} \Delta^{\mathbf{k}\mathbf{k},\mathbf{q}\mathbf{q}}_{\alpha \beta} \hat{c}^{\dagger}_{e,\mathbf{k}} \hat{c}^{\dagger}_{g_{\beta},\mathbf{q}} \hat{c}_{g_{\alpha},\mathbf{k}} \hat{c}_{e,\mathbf{q}} \notag\\
    &\qquad \quad  + \sum_{\mathbf{k} \neq \mathbf{q}} \Delta^{\mathbf{k}\mathbf{q},\mathbf{q}\mathbf{k}}_{\alpha \beta} \hat{c}^{\dagger}_{e,\mathbf{k}} \hat{c}^{\dagger}_{g_{\beta},\mathbf{q}} \hat{c}_{g_{\alpha},\mathbf{q}} \hat{c}_{e,\mathbf{k}} \bigg),\label{eq:H_ME} \\
      \mathcal{L}(\hat{\rho}) &=  \sum_{\alpha, \beta}\bigg( \sum_{\mathbf{k},\mathbf{q}} \Gamma^{\mathbf{k}\mathbf{k},\mathbf{q}\mathbf{q}}_{\alpha \beta} \left(  2 \hat{\sigma}^{\mathbf{q}\mathbf{q}}_{g_{\beta} e} \hat{\rho} \hat{\sigma}^{\mathbf{k}\mathbf{k}}_{e g_{\alpha}} - \left\{\hat{\sigma}^{\mathbf{k}\mathbf{k}}_{eg_{\alpha}} \hat{\sigma}^{\mathbf{q}\mathbf{q}}_{g_{\beta} e}, \hat{\rho}  \right\} \right)\notag \\
    &\quad+ \sum_{\mathbf{k} \neq \mathbf{q}} \Gamma^{\mathbf{k}\mathbf{q},\mathbf{q}\mathbf{k}}_{\alpha\beta} \left(  2 \hat{\sigma}^{\mathbf{q}\mathbf{k}}_{g_{\beta} e} \hat{\rho} \hat{\sigma}^{\mathbf{k}\mathbf{q}}_{eg_{\alpha}} -  \left\{\hat{\sigma}^{\mathbf{k}\mathbf{q}}_{e g_{\alpha}} \hat{\sigma}^{\mathbf{q}\mathbf{k}}_{g_{\beta} e}, \hat{\rho}  \right\} \right)\bigg). \label{eq:L_ME} 
\end{align}
Here, $\hat{\sigma}_{e, g_{\alpha}}^{\mathbf{k}\mathbf{q}} = \hat{c}^{\dagger}_{e \mathbf{k}} \hat{c}_{g_{\alpha},\mathbf{q}}$, and $\hat{c}^{\dagger}_{e,\mathbf{k}}$($\hat{c}^{\dagger}_{g_{\alpha},\mathbf{k}}$) creates a fermion in the $e$ ($g_{\alpha}$) state with in-plane momentum $\mathbf{k}=(k_x,k_y)$ in the harmonic oscillator ground state $n_{0,z}$ along $z$.

The terms $\Delta_{\alpha \beta}^{\mathbf{kl,mn}}$ ($\Gamma_{\alpha \beta}^{\mathbf{kl,mn}})$ describe coherent (incoherent) exchange of photons of the relevant transitions $\alpha,\beta=0,1$ between two atoms in the corresponding internal and motional states (Fig.~\ref{fig:Fig1}d,e). They are defined as  projections of the real (R) and imaginary part (I) of the Green's tensor $G$ as $\Delta^{\mathbf{ij,kl}}_{\alpha \beta} = \mathbf{d}_{\alpha}^T G^{\mathbf{ij,kl}}_R \bar{\mathbf{d}}_{\beta} $, $\Gamma^{\mathbf{ij,kl}}_{\alpha\beta} = \mathbf{d}_{\alpha}^T G_I^{\mathbf{ij,kl}} \bar{\mathbf{d}}_{\beta}$, where $\mathbf{d}_{\alpha} = C_{\alpha} \mathbf{n}_{\alpha}$, given in terms of the Clebsch-Gordan coefficient $C_{\alpha}$ and the polarization vector $\mathbf{n}_{\alpha}$ ($\mathbf{n}_0 = \mathbf{e}_x$, $\mathbf{n}_1 =(\mathbf{e}_y+i\mathbf{e}_z)/\sqrt{2}$), and $\mathbf{d}^T$, $\bar{\mathbf{d}}$ denote the transpose and complex conjugate, respectively. The Green's tensor is $G(r) = \frac{3 \Gamma}{4} \left\{\left[\mathbf{I}-\hat{\mathbf{r}} \otimes \hat{\mathbf{r}} \right] \frac{e^{i k_0 r}}{k_0 r}  + \left[\mathbf{I}-3\hat{\mathbf{r}} \otimes \hat{\mathbf{r}} \right] \left[\frac{i e^{i k_0 r}}{(k_0 r)^2} - \frac{e^{i k_0 r}}{(k_0 r)^3} \right]\right\}$, with the wavevector $k_0$ of the ground-excited state transition. The matrix elements $G^{\mathbf{ij,kl}}$ are defined as $  G^{\mathbf{ij,kl}} = \int d\mathbf{r} d\mathbf{r}^{\prime}  \, \bar{\phi}_{\mathbf{i}}(\mathbf{r}) \phi_{\mathbf{j}}(\mathbf{r}) \, G(\mathbf{r} - \mathbf{r}^{\prime}) \,  \bar{\phi}_{\mathbf{k}}(\mathbf{r}^{\prime}) \phi_{\mathbf{l}}(\mathbf{r}^{\prime})$, where  $\phi_{\mathbf{k}}(x,y,z) = \frac{1}{\sqrt{A}}  e^{i(k_x x+k_y y)} \psi_0(z) $. The area $A$ is chosen to approximate a harmonically trapped gas with trapping frequency $\omega_{\perp}= 2\pi \times 150$ Hz \cite{supplemental}.

 We then derive equations of motion by using a mean field approximation, which   factorizes 4-operator terms as products of 2-operator terms (see \cite{supplemental}). Given the uncorrelated  initial conditions this treatment  is well justified at short times.  We further assume that the dynamics is dominated by the momentum diagonal elements of the density matrix, $\rho^{\mu\nu}_{\mathbf{q}\mathbf{q}} = \left< \hat{c}^{\dagger}_{\mu,\mathbf{q}} \hat{c}_{\nu,\mathbf{q}}\right>$, where $\mu,\nu = e,g_0,g_1$, given  the lack of momentum off-diagonal coherences in the initial state. Under these approximations the  population of the excited state in momentum $\mathbf{q}$ evolves as
\begin{align}
 \frac{d \rho^{ee}_{\mathbf{q}\mathbf{q}}}{dt} &= \sum_{\alpha} \sum_k -2 \Gamma^{\mathbf{k}\mathbf{q},\mathbf{q}\mathbf{k}}_{\alpha\alpha} (1-\rho^{g_{\alpha}g_{\alpha}}_{\mathbf{k}\mathbf{k}}) \rho^{ee}_{\mathbf{q}\mathbf{q}} \label{eq:ME_semiclassical_part}\\
  & + \sum_{\alpha,\beta} \sum_k i \left( \mathcal{G}^{\mathbf{qq,kk}}_{\alpha \beta} \rho^{eg_{\alpha}}_{\mathbf{q}\mathbf{q}} \rho^{g_{\beta}e}_{\mathbf{k}\mathbf{k}}   -\bar{\mathcal{G}}^{\mathbf{qq,kk}}_{\alpha \beta}\rho^{g_{\beta}e}_{\mathbf{q}\mathbf{q}}  \rho^{eg_{\alpha}}_{\mathbf{k}\mathbf{k}}  \right)\notag ,
\end{align}
where $\mathcal{G}^{\mathbf{k}\mathbf{k},\mathbf{q}\mathbf{q}}_{\alpha \beta} =\Delta^{\mathbf{k}\mathbf{k},\mathbf{q}\mathbf{q}}_{\alpha \beta} +i \Gamma^{\mathbf{k}\mathbf{k},\mathbf{q}\mathbf{q}}_{\alpha \beta} $.  The first line corresponds to the spontaneous decay process  $ \ket{e,\mathbf{q},n_{0,z}} \to  \ket{g_\alpha,\mathbf{k},n_{0,z}} $ at a rate set by  $\Gamma_{\alpha\alpha}^{\mathbf{q}\mathbf{k},\mathbf{k}\mathbf{q}}$, which accounts for the momentum conservation in the emission process, and is Pauli suppressed by the factor $1-\rho^{g_{\alpha}g_{\alpha}}_{\mathbf{k}\mathbf{k}}$. Thus, the ME in momentum space naturally recovers Pauli blocking for arbitrary  multilevel structures and generic geometries, while also including  the  most  relevant  cooperative  effects, such as superradiance and subradiance, that emerge from the terms in the second line. The latter depend on the coherences, $\rho^{e g_{\alpha}}$, between the excited and the ground state atoms and have contributions from the coherent ($\Delta$) and incoherent dipolar exchange processes  ($\Gamma$).

\begin{figure}
\includegraphics[width=\columnwidth]{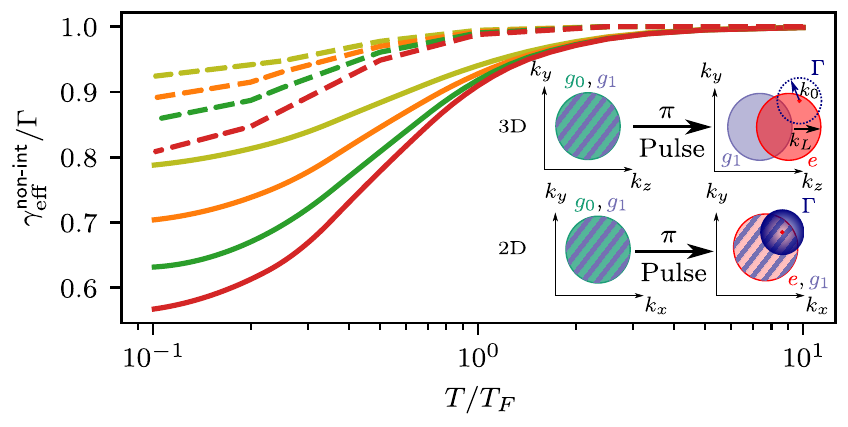}
\caption{Pauli blocking of spontaneous emission for non-interacting atoms.  Decay rate $\gamma_{\mathrm{eff}}^{\text{non-int}}/\Gamma$ vs $T/T_F$ comparing 3D (dashed) to 2D (solid) for  $N_0=N_1$, and  $\theta=\pi$.  $N_{2D}=100,\,200,\,400,\,800$ for the colored lines (top to bottom), corresponding to $k_0/k_F = 1.51,\, 1.26,\, 1.07,\, 0.90 $. $N_{3D}$ is chosen such that ${ k}_0/k_F^{3D} = { k}_0/k_F^{2D}$ at these atom numbers. Inset: Illustration of the initial state before and after laser excitation in 2D and 3D. Colored circles denote Fermi seas of different atomic levels and striped areas denote an overlap of two Fermi seas. Blue circles of radius ${ k}_0$ labelled $\Gamma$ show the states reachable by a spontaneous decay process.%
\label{fig:Fig2}}
\end{figure}
{\it Pauli blocking in non-interacting atoms.---}We  start by studying  the first line of Eq.~(\ref{eq:ME_semiclassical_part}) fully neglecting interaction effects, and note the close resemblance to prior semi-classical approaches \cite{Shuve_2009} where the decay of the atoms is dictated solely by the volume of the available phase-space. To gain intuition note that $\Gamma_{\alpha\alpha}^{\bf \mathbf{q}\mathbf{k},\mathbf{k}\mathbf{q}}$ mediates the decay of an atom  $e$ with momentum $\mathbf{q}$ to the ground state $g_{\alpha}$ at momentum $\mathbf{k}$ if $|\mathbf{k} - \mathbf{q}| \le k_0$ (2D) or $|\mathbf{k} - \mathbf{q}| = k_0$ (3D)~\cite{supplemental}. This decay will be Pauli blocked if the corresponding state is occupied due to the factor $1-\rho^{g_{\alpha}g_{\alpha}}_{\mathbf{k}\mathbf{k}}$. Consequently, in the presence of a Fermi sea of ground-state atoms, the decay rate will be reduced. The degree of Pauli blocking will depend on  the ratio of $k_0$ to $k_F$ (controlled by  the density), and the mean occupation within the Fermi sea (set by the temperature of the gas).

To illustrate this phenomenology we show in Fig.~\ref{fig:Fig2} the rate $\gamma_\text{eff}^\text{non-int}/\Gamma$ for an initially balanced Fermi gas  with  $N_0=N_1$ non-interacting atoms, and full excitation of all $g_0$ atoms to the $e$ state (i.e.~$\theta=\pi$), as a function of the temperature $T/T_F$. In this case the maximal Pauli suppression achievable is  $\Gamma_{11}/\Gamma \sim 81 \%$  when  all decay channels into $g_1$ are blocked. We compare the 3D (dashed) to the strongly confined 2D system (solid) for a range of $N$ (colors), i.e.~of $ k_0/k_F$. Most notably, we observe a strong enhancement of Pauli blocking in 2D compared to 3D, over a significantly larger range of temperatures.

This can be understood as follows. Firstly, the axial confinement changes the energy spectrum, and thus the density of states. This results in a higher mean occupation fraction in 2D, and consequently stronger Pauli blocking than in 3D for the same $k_0/k_F$. Secondly, the initial laser excitation imparts a momentum kick to the atoms in 3D that displaces the excited population  away from the unexcited ground state Fermi sea facilitating decay to unoccupied states (inset of Fig.~\ref{fig:Fig2}). In contrast, in 2D for laser excitation along the strongly confined direction, the motional states are unaffected enhancing the probability to decay to an already occupied state.

{\it Interplay of dipolar interactions  with Pauli blocking.---}%
\begin{figure}
\includegraphics[width=\columnwidth]{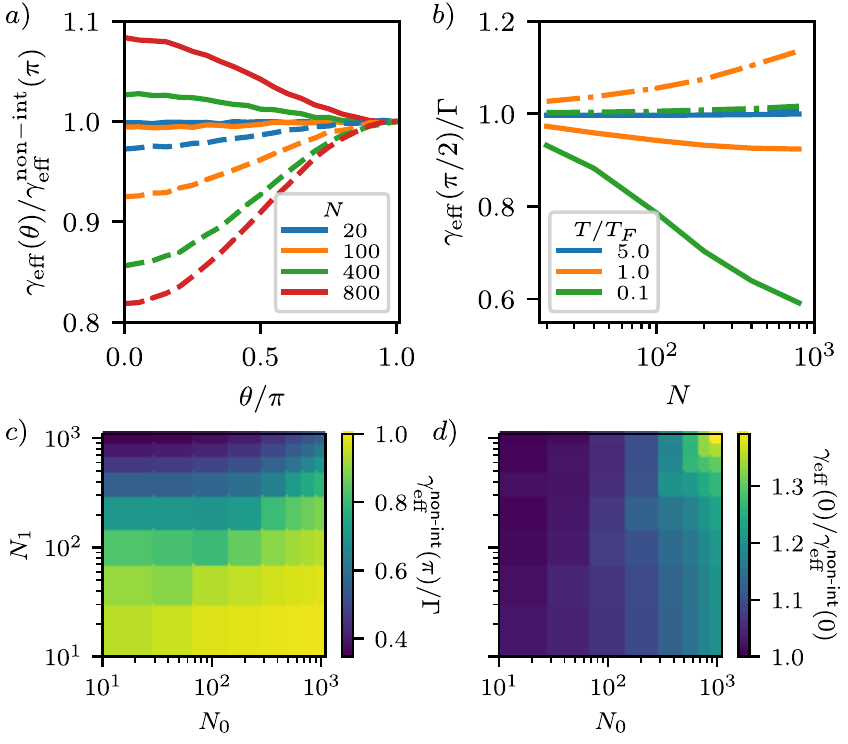}
\caption{Interplay of dipolar interactions and Pauli blocking in 2D. a)  $\gamma_{\mathrm{eff}}(\theta)/\gamma_{\mathrm{eff}}^{\text{non-int}}(\pi)$ comparing the full ME (solid) to the non-interacting  part (dashed) at $T/T_F=0.1$ for different $N=N_0=N_1$. b) Decay rate $\gamma_{\mathrm{eff}}/\Gamma$ at $\theta=\pi/2$ versus $N$ at different temperatures comparing the full ME (solid) to the frozen atom approximation (dashed-dotted). c) Decay rate $\gamma^{\mathrm{non-int}}_{\mathrm{eff}}(\pi)$ versus $N_0$ and $N_1$. d) Interaction effects quantified by $\gamma_{\mathrm{eff}}/\gamma^{\mathrm{non-int}}_{\mathrm{eff}}$ for $\theta \rightarrow 0$ versus $N_0$, $N_1$. c) and d) are in thermal equilibrium for $T_0=T_1$ and $T/T_{F,1}=0.1 $.\label{fig:Fig3}
}
\end{figure} 
In Fig.~\ref{fig:Fig3} we study how $\gamma_\text{eff}^\text{non-int}$ (dashed lines) is modified by dipolar induced cooperative effects  using the full master equation (ME) (solid lines)  as a function of the pulse area  $\theta$  for a balanced gas, $N_0=N_1$.  At $\theta=\pi$, interactions have no effect on $\gamma_{\mathrm{eff}}$ due to the absence of initial coherences.
However, for smaller $\theta$ and increasing $N$ there is an intricate competition between interactions and Pauli blocking. Note the  modifications from  $\theta$ affect exclusively the $e\rightarrow g_0$ transition. On the one hand, lowering $\theta$ results in a higher population in the $g_0$ ground state $\sim \cos^2(\theta/2)$ which   increases  Pauli blocking of the $e\rightarrow g_0$ decay channel for non-interacting atoms. On the other hand,  interaction effects become  stronger at low $\theta$, leading to an enhanced normalized superradiant decay  which scales as $\sim \cos^2(\theta/2)$~\cite{supplemental}. Importantly, the superradiant enhancement also scales with $N$. This interplay leads to  dominant Pauli blocking  and thus lower decay rates  at low atom numbers and dominant cooperatively enhanced emission and faster radiative decay  at high densities as $\theta$ decreases.

To demonstrate the importance of including atomic motion  and its interplay with quantum statistics  we  also compare the ME in momentum space to the usual frozen atom approximation (FA) (\cite{PhysRevA.94.023612,PhysRevA.2.883,PhysRevA.47.1336,PhysRevA.55.513,PhysRevA.81.053821,JoMO_Kaiser_2011,PhysRevLett.108.123602,PhysRevLett.116.083601,PhysRevA.93.023407,Bromley2016,Weiss_2018,PhysRevLett.117.073003,PhysRevLett.116.083601}), which is derived for atoms assumed to be at fixed positions in real space. We show the resulting $\gamma_{\mathrm{eff}}$ for a $\theta=\pi/2$ excitation as a function of the atom number $N$ and temperature $T/T_F$ in Fig.~\ref{fig:Fig3}b. The FA properly captures the superradiantly enhanced decay rate due to dipolar interactions. However, its inability to account for  Pauli blocking results in incorrect predictions in the quantum degenerate regime, and an incorrect scaling of the decay rate with  $N$.

Finally, Figs.~\ref{fig:Fig3}c and \ref{fig:Fig3}d explore the role of the imbalance $N_1/N_0$ of the initial populations of the two ground  states on the effective decay rate. For non-interacting atoms  it is highly advantageous to only excite a small fraction of atoms  (\cite{Shuve_2009}) to  maximize Pauli blocking. This is demonstrated in panel c which shows the decay rate $\gamma_{\mathrm{eff}}^{\mathrm{non-int}}/\Gamma$ at $T/T_F=0.1$ for $\theta=\pi$, predicting the largest suppression in the $N_1 \gg N_0$ regime. Interestingly, for  the 2D  system  even in the presence of superradiance as  $\theta \rightarrow 0$  it is  possible to  minimize interaction effects while maintaining significant Pauli suppression ($\sim 66 \%$) by choosing  $N_1 \gg N_0$ as shown in \ref{fig:Fig3}d. This is because  only the $N_0$ atoms feature coherences and experience dipolar interactions at early times.

{\it Comparison with  experiment.---} 
\begin{figure}
\includegraphics[width=\columnwidth]{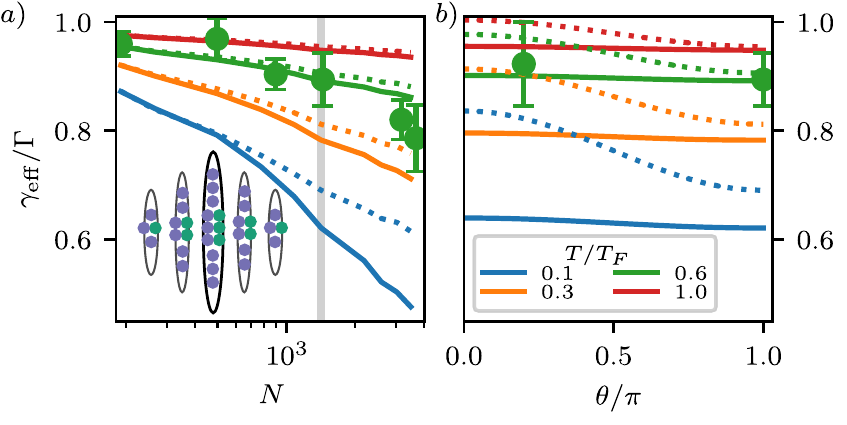}
\caption{Decay rates in stacked pancakes geometry. a) Scaling of the decay rate $\gamma_{\mathrm{eff}}$ with $N$ comparing the balanced case (dotted), $N_0=N_1=N$,  to the highly-imbalanced case (solid), $N_0=200,N_1=N$, for a $\theta=\pi$ pulse. The gray shading denotes the fixed $N=1500$ used in panel b)  $\gamma_{\mathrm{eff}}/\Gamma$ vs  $\theta$. Temperatures indicated in the legend in b, in the imbalanced case $T_0=T_1$ and $T/T_F$ given with respect to $T_{F,1}$. Points with error bars are experimental data taken under the same  conditions as the solid green lines.
\label{fig:Fig4}}
\end{figure}
Here we perform ME simulations for an array of two-dimensional pancakes (inset of Fig.~\ref{fig:Fig4}), realised by confining the initially 3D Fermi gas in a deep optical lattice along $z$,  and compare with experimental observations. The extension of the theory, and details on the experimental preparation and measurement protocol are provided in \cite{supplemental}.

Fig.~\ref{fig:Fig4}a shows the effective  decay rate $\gamma_{\mathrm{eff}}/\Gamma $ as a function of the atom number, pulse area  and temperature, comparing a balanced (dotted) to the imbalanced case (solid) for a $\pi$ pulse excitation. For $N \sim 4 \times 10^3$ and  $T/T_F = 0.1$ corresponding to peak densities of $10^{14} \, \mathrm{cm}^{-3}$,  we predict up to a factor of 2 enhancement in the atomic lifetime. In Fig.~\ref{fig:Fig4}b we demonstrate the strong interaction effects present for balanced gases when preparing initial coherences due to superradiance. In contrast, we emphasise the weak dependence on $\theta$ in the imbalanced case, which demonstrates the lack of significant cooperative effects.

In Fig.~\ref{fig:Fig4} we also  include  experimental data available for the highly imbalanced scenario,  which appears to be consistent with our theory predictions within errors (green points and lines). The excellent agreement with the ME, in a regime where interactions are shown to only weakly modify the decay, supports the validity of the theoretical model and motivates further experimental work to demonstrate Pauli blocking in population measurements. A particularly important future direction is to prepare a deeply degenerate Fermi gas in 2D, where the differences between balanced and imbalanced cases become rather striking.

{\it Outlook.---} We identified a regime in imbalanced multilevel 2D Fermi gases where superradiant effects are weak and thus enhanced lifetimes attributable to Pauli blocking can be directly observed via population measurements. While our calculations are restricted to short times where a mean-field analysis is valid, as confirmed by comparisons with experimental measurements, they will break down at longer times. This in turn might give rise to genuine many-body effects and unexpected novel behavior yet to be explored. 

 \begin{acknowledgments}
\noindent{\textit{Acknowledgements:}} 
AFOSR Grant No. FA9550-18-1-0319 and its MURI Initiative, by the DARPA and ARO Grant No. W911NF-16-1-0576, the ARO single investigator Grant No. W911NF-19-1-0210, the NSF PHY1820885, NSF JILA-PFC PHY-1734006 Grants, NSF QLCI-2016244 grant, DOE-QSA and by NIST. C.S. thanks the Humboldt Foundation for support. The authors thank Nathan Schine and Joseph Thywissen for providing feedback on the manuscript.
\end{acknowledgments}



\bibliography{pauli_blocking}{}

\begin{appendix}
  
 \begin{center}
   \textbf{\large Supplemental Material}
 \end{center}
\setcounter{equation}{0}
\setcounter{figure}{0}
\setcounter{table}{0}
\makeatletter
\renewcommand{\theequation}{S\arabic{equation}}
\renewcommand{\thefigure}{S\arabic{figure}}
\renewcommand{\thetable}{S\arabic{table}}
\setcounter{section}{0}
\renewcommand{\thesection}{S-\Roman{section}}


\subsection{Level structure and geometry}
We consider a two-dimensional system engineered by tightly  confining  a gas of ultracold fermionic atoms along one direction, $z$, and applying only a weak  in-plane confinement along $x$ and $y$.
We assume the system is in the regime where only the ground state harmonic oscillator mode $n_{0,z}$ along $z$ is occupied, but motion is allowed in $x$ and $y$.

Moreover, each atom has a $\Lambda$-type internal electronic level structure, where an excited state $e$ can spontaneously decay into the ground states $g_0$ or $g_1$ with decay rates $\Gamma_{00}$ or $\Gamma_{11}$, respectively.
Specficially, we consider the ${}^1S_0\, (F=9/2)$ to $ {}^3P_1\, (F=11/2)$ transition in ${}^{87}\text{Sr}$.
Atoms are prepared in the $m_F=-9/2$ and $m_F=-7/2$ ground states, and optically excited to the $m_F=-9/2$ state in the excited state manifold.
Using the notation $\ket{F,m_F}$ for angular momentum states, we label the two ground states as $g_0 = \ket{9/2,-9/2}$, $g_1 = \ket{9/2,-7/2}$, and the excited state as $e = \ket{11/2,-9/2}$. We will assume that the emitted photon for $e\rightarrow g_0\, (g_1)$ has $\pi$ ($\sigma^-$) polarization with quantization axis along $x$.

In principle, dipolar exchange interactions can couple the $g_0-e$ and $g_1-e$ transitions to other internal transitions, and lead to population transfer to other Zeeman sublevels.
However, by applying a strong magnetic field the $m_F$ levels are split in energy, making such processes off-resonant, and thus fully suppressed.
In this way, we can limit our internal states to a 3 level $\Lambda$ subsystem.

\subsection{Dipolar multilevel master equation}
The dipolar multilevel master equation derived in Ref.~\cite{PhysRevA.101.043816}, restricted to the $\Lambda$ level configuration described above, is given by $\dot{\hat{\rho}} = -i \left[\hat{H},\hat{\rho} \right] + \mathcal{L}(\hat{\rho})$, with 
\begin{align}
    \hat{H} &= - \sum_{\alpha,\beta} \int d\vect{r} d\vect{r}^{\prime}   \left(\vect{d}_{\alpha} \cdot \mathrm{Re}\,G(\vect{r}-\vect{r}^{\prime})  \cdot \vect{\bar{d}}_{\beta}\right) \notag \\
    & \qquad \qquad \qquad \hat{\sigma}_{e g_{\alpha}}(\vect{r}) \, \hat{\sigma}_{g_{\beta} e}(\vect{r}^{\prime}),
    \label{eq_sup:H_full}\\
    \mathcal{L}(\hat{\rho}) &= - \sum_{\alpha,\beta} \int d\vect{r} d\vect{r}^{\prime}  \left(\vect{d}_{\alpha} \cdot \mathrm{Im}\,G(\vect{r}-\vect{r}^{\prime})  \cdot \vect{\bar{d}}_{\beta}\right) \notag\\
    & \left(  \left\{\hat{\sigma}_{eg_{\alpha}}(\vect{r}) \hat{\sigma}_{g_{\beta} e}(\vect{r}^{\prime}), \hat{\rho}  \right\} - 2 \hat{\sigma}_{g_{\beta} e}(\vect{r}^{\prime})\hat{\rho} \hat{\sigma}_{eg_{\alpha}}(\vect{r}) \right),
    \label{eq_sup:L_full}
\end{align}
where $\hat{\sigma}_{g_{\alpha} e}(\vect{r}) = \hat{c}^{\dagger}_{g_{\alpha}}(\vect{r}) \hat{c}_e(\vect{r})$ destroys a fermionic atom in excited state $e$ at positon $\vect{r}$ and creates a fermionic atom in ground state $g_{\alpha}$ at positon $\vect{r}$. The operators $\hat{c}^{(\dagger)}_{\lambda}(\vect{r})$, $\lambda\in\{g_0,g_1,e\}$, fulfill the usual fermion anticommutation relations.

The electromagnetic Green's tensor at position $\mathbf{r}$ is given by
\begin{align}
    G(\mathbf{r}) &= \frac{3 \Gamma}{4} \left\{\left[\mathbf{I}-\hat{\mathbf{r}} \otimes \hat{\mathbf{r}} \right] \frac{e^{i k_0 r}}{k_0 r} \right. \nonumber \\
    & \quad \quad \quad +\left. \left[\mathbf{I}-3\hat{\mathbf{r}} \otimes \hat{\mathbf{r}} \right] \left[\frac{i e^{i k_0 r}}{(k_0 r)^2} - \frac{e^{i k_0 r}}{(k_0 r)^3} \right]\right\},
\end{align}
where $\hat{\mathbf{r}}=\mathbf{r}/r$, $r\equiv|\mathbf{r}|$, and $\mathbf{I}$ is a three-component identity matrix.
The dipole operators are $\mathbf{d}_{\alpha} = C_{\alpha} \mathbf{n}_{\alpha}$ which depend on the Clebsch-Gordon coefficient $C_{\alpha} = \left<9/2, -9/2+\alpha; 1, -\alpha| 11/2,-9/2\right>$
and the polarisation vector $\mathbf{n}_{\alpha}$, $\mathbf{n}_0 = \mathbf{e}_x$, $\mathbf{n}_1 =(\mathbf{e}_y+i\mathbf{e}_z)/\sqrt{2}$, of the relevant transitions.

\subsection{Lindbladian in momentum space}
Expanding the atomic creation operators in a single particle basis as $\hat{c}^{\dagger}(\vect{r}) = \sum_{\mathbf{i}} \bar{\phi}_{\mathbf{i}}(\vect{r}) \hat{c}^{\dagger}_{\mathbf{i}}$, where $\hat{c}^{\dagger}_{\mathbf{i}}$ creates a fermionic atom in the state described by the wave-function $\phi_{\mathbf{i}}(\vect{r})$,
we obtain from Eqs.~(\ref{eq_sup:H_full}) and (\ref{eq_sup:L_full}) the model
\begin{align}
    \hat{H} &=  \sum_{\alpha, \beta} \sum_{\mathbf{\mathbf{ijkl}}} \Delta^{\mathbf{ij,kl}}_{\alpha \beta} \hat{c}^{\dagger}_{e,\mathbf{i}} \hat{c}^{\dagger}_{g_{\beta},\mathbf{k}} \hat{c}_{g_{\alpha},\mathbf{j}} \hat{c}_{e,\mathbf{l}}, \label{eq_sup:Hdd} \\
    \mathcal{L}(\hat{\rho}) &= - \sum_{\alpha, \beta} \sum_{\mathbf{ijkl}} \Gamma^{\mathbf{ij,kl}}_{\alpha \beta} \left(  \left\{\hat{\sigma}^{\mathbf{ij}}_{eg_{\alpha}} \hat{\sigma}^{\mathbf{kl}}_{g_{\beta} e}, \hat{\rho}  \right\} - 2 \hat{\sigma}^{\mathbf{kl}}_{g_{\beta} e} \hat{\rho} \hat{\sigma}^{\mathbf{ij}}_{e g_{\alpha}} \right),\label{eq_sup:L} 
\end{align}
where we normal ordered the Hamiltonian part to not introduce spurious self-interactions and the matrix elements are defined as
\begin{align}
    \Delta^{\mathbf{ij,kl}}_{\alpha\beta} &= \int d\mathbf{r} d\mathbf{r}^{\prime}  \, \bar{\phi}_{\mathbf{i}}(\mathbf{r}) \phi_{\mathbf{j}}(\mathbf{r}) \, \mathrm{Re} \, G_{\alpha\beta}(\mathbf{r} - \mathbf{r}^{\prime}) \,  \bar{\phi}_{\mathbf{k}}(\mathbf{r}^{\prime}) \phi_{\mathbf{l}}(\mathbf{r}^{\prime}),
    \label{eq_sup:Delta_ijkl} \\
    \Gamma^{\mathbf{ij,kl}}_{\alpha\beta} &= \int d\mathbf{r} d\mathbf{r}^{\prime}  \, \bar{\phi}_{\mathbf{i}}(\mathbf{r}) \phi_{\mathbf{j}}(\mathbf{r}) \, \mathrm{Im} \, G_{\alpha\beta}(\mathbf{r} - \mathbf{r}^{\prime}) \,  \bar{\phi}_{\mathbf{k}}(\mathbf{r}^{\prime}) \phi_{\mathbf{l}}(\mathbf{r}^{\prime}),
    \label{eq_sup:Gamma_ijkl}
\end{align}
with the projections
\begin{align}
    G_{00} &= C_{0}^2 \, \left(\vect{n}_0  \cdot G \cdot \bar{\vect{n}}_0\right),\\
    G_{11} &=C_{1}^2 \,\left( \vect{n}_1 \cdot G\cdot  \bar{\vect{n}}_1 \right),\\
    G_{01} &=     \bar{G}_{10}=C_{0} C_{1} \, \left( \vect{n}_0\cdot  G\cdot  \bar{\vect{n}}_{1}\right).
\end{align}

\subsection{Mode-conserving approximation}
In a first approximation we only keep terms  that conserve the modes of the involved particles, e.g. we match the creation and annihilation operators in Eqs.~(\ref{eq_sup:Hdd}) and (\ref{eq_sup:L}) by setting $i=j$, $k=l$ or $i=l$, $k=j$. 
While this approximation is uncontrolled it has been shown to be effective in describing the dynamics of a variety of systems (citations). 

This results in the master equation presented in the main text, 
\begin{align}
    \hat{H} &=  \sum_{\alpha, \beta} \left(\sum_{\mathbf{k}\mathbf{q}} \Delta^{\mathbf{k}\mathbf{k},\mathbf{q}\mathbf{q}}_{\alpha \beta} \hat{c}^{\dagger}_{e,\mathbf{k}} \hat{c}^{\dagger}_{g_{\beta},\mathbf{q}} \hat{c}_{g_{\alpha},\mathbf{k}} \hat{c}_{e,\mathbf{q}} \right.\nonumber\\
    &\qquad \qquad  + \left.\sum_{\mathbf{k} \neq \mathbf{q}} \Delta^{\mathbf{k}\mathbf{q},\mathbf{q}\mathbf{k}}_{\alpha \beta} \hat{c}^{\dagger}_{e,\mathbf{k}} \hat{c}^{\dagger}_{g_{\beta},\mathbf{q}} \hat{c}_{g_{\alpha},\mathbf{q}} \hat{c}_{e,\mathbf{k}} \right),
    \label{eq_sup:H_final}\\
    \mathcal{L}(\hat{\rho}) &= \sum_{\alpha, \beta}\left( \sum_{\mathbf{k}\mathbf{q}} \Gamma^{\mathbf{k}\mathbf{k},\mathbf{q}\mathbf{q}}_{\alpha \beta} \left( 2 \hat{\sigma}^{\mathbf{q}\mathbf{q}}_{g_{\beta} e} \hat{\rho} \hat{\sigma}^{\mathbf{k}\mathbf{k}}_{e g_{\alpha}} -\left\{\hat{\sigma}^{\mathbf{k}\mathbf{k}}_{eg_{\alpha}} \hat{\sigma}^{\mathbf{q}\mathbf{q}}_{g_{\beta} e}, \hat{\rho}  \right\}\right) \right. \nonumber\\
    &\quad  +\left. \sum_{\mathbf{k} \neq \mathbf{q}} \Gamma^{\mathbf{k}\mathbf{q},\mathbf{q}\mathbf{k}}_{\alpha\beta} \left(2 \hat{\sigma}^{\mathbf{q}\mathbf{k}}_{g_{\beta} e} \hat{\rho} \hat{\sigma}^{\mathbf{k}\mathbf{q}}_{eg_{\alpha}} -\left\{\hat{\sigma}^{\mathbf{k}\mathbf{q}}_{e g_{\alpha}} \hat{\sigma}^{\mathbf{q}\mathbf{k}}_{g_\beta e}, \hat{\rho}  \right\}\right)\right).
    \label{eq_sup:L_final}
\end{align}

\subsection{Mean field equations of motion}
Starting from the master equation of Eqs.~(\ref{eq_sup:H_final}) and (\ref{eq_sup:L_final}) we can derive equations of motion for the elements of the density matrix. 
Since we used a mode-conserving approximation, we also only consider the evolution of mode-diagonal elements of the density matrix, specifically
$\rho^{\mu\nu}_{\mathbf{q}\mathbf{q}} = \left< \hat{c}^{\dagger}_{\mu,\mathbf{q}} \hat{c}_{\nu,\mathbf{q}}\right>$ with $\mu,\nu=e,g_0,g_1$.

When deriving equations of motion for these two-body operators we generically encounter expectation values of 4-body operators of the form $ \sexpval{\hat{c}^{\dagger}_{\mu,\mathbf{i}}\hat{c}^{\dagger}_{\nu,\mathbf{j}}\hat{c}_{\mu^{\prime},\mathbf{k}}\hat{c}_{\nu^{\prime},\mathbf{l}}}$.
We factorise these as
\begin{align*}
 \sexpval{\hat{c}^{\dagger}_{\mu,\mathbf{i}}\hat{c}^{\dagger}_{\nu,\mathbf{j}}\hat{c}_{\mu^{\prime},\mathbf{k}}\hat{c}_{\nu^{\prime},\mathbf{l}}}&\approx \sexpval{\hat{c}^{\dagger}_{\mu,\mathbf{i}}\hat{c}_{\nu^{\prime},\mathbf{l}}}\sexpval{\hat{c}^{\dagger}_{\nu,\mathbf{j}}\hat{c}_{\mu^{\prime},\mathbf{k}}} -\sexpval{\hat{c}^{\dagger}_{\mu,\mathbf{i}}\hat{c}_{\mu^{\prime},\mathbf{k}} }\sexpval{\hat{c}^{\dagger}_{\nu,\mathbf{j}}\hat{c}_{\nu^{\prime},\mathbf{l}}}\\
 &= \sexpval{\hat{\sigma}^{\mu\nu^{\prime}}_{\mathbf{il}}} \sexpval{\hat{\sigma}^{\nu\mu^{\prime}}_{\mathbf{jk}}}-\sexpval{\hat{\sigma}^{\mu\mu^{\prime}}_{\mathbf{ik}}} \sexpval{\hat{\sigma}^{\nu\nu^{\prime}}_{\mathbf{jl}}}\\
 &\approx \delta_{\mathbf{il}} \delta_{\mathbf{jk}} \sexpval{\hat{\sigma}^{\mu\nu^{\prime}}_{\mathbf{il}}} \sexpval{\hat{\sigma}^{\nu\mu^{\prime}}_{\mathbf{jk}}}-\delta_{\mathbf{ik}} \delta_{\mathbf{jl}}\sexpval{\hat{\sigma}^{\mu\mu^{\prime}}_{\mathbf{ik}}} \sexpval{\hat{\sigma}^{\nu\nu^{\prime}}_{\mathbf{jl}}},
\end{align*}
where the first approximation assumes that 4-body operator factorise into 2-body operators as for a non-interacting Fermi gas, and the second approximation assumes that there are no mode-off-diagonal correlations. The first approximation is justified as long as there are no strong correlations in the initial state, and as long as interactions do not result in significant correlations during time-evolution. The second approximation is exact for the initial state we consider, but will become invalid as coherences between different momenta build up during time evolution. However, since the initial dynamics will be dominated by the initial coherences which are diagonal, we expect this latter approximation to be good at least for short times.

With these approximations we then obtain the equations of motion as

\begin{equation}
\begin{split}
    \frac{d \rho^{ee}_{\mathbf{q}\mathbf{q}}}{dt} &=\sum_{\alpha} \sum_k -2 (1-\rho^{g_{\alpha}g_{\alpha}}_{\mathbf{k}\mathbf{k}}) \rho^{ee}_{\mathbf{q}\mathbf{q}} \Gamma^{\mathbf{k}\mathbf{q},\mathbf{q}\mathbf{k}}_{\alpha\alpha} \\
  & \quad + i \sum_{\alpha,\beta} \sum_k  \left(\rho^{eg_{\alpha}}_{\mathbf{q}\mathbf{q}} \rho^{g_{\beta}e}_{\mathbf{k}\mathbf{k}}  \mathcal{G}^{\mathbf{qq,kk}}_{\alpha \beta}  -\rho^{g_{\beta}e}_{\mathbf{q}\mathbf{q}}  \rho^{eg_{\alpha}}_{\mathbf{k}\mathbf{k}} \bar{\mathcal{G}}^{\mathbf{qq,kk}}_{\alpha \beta} \right) 
\end{split}
\label{eq_sup:rhoee}
\end{equation}

\begin{equation}
\begin{split}
  \frac{d \rho^{g_{\alpha}g_{\alpha}}_{\mathbf{q}\mathbf{q}}}{dt} &= \sum_k 2 (1-\rho^{g_{\alpha}g_{\alpha}}_{\mathbf{q}\mathbf{q}}) \rho^{ee}_{\mathbf{k}\mathbf{k}} \Gamma^{\mathbf{k}\mathbf{q},\mathbf{q}\mathbf{k}}_{\beta\beta}  \\
  & -i\sum_{\beta} \sum_k  \left(\rho^{eg_{\alpha}}_{\mathbf{q}\mathbf{q}} \rho^{g_{\beta}e}_{\mathbf{k}\mathbf{k}}  \mathcal{G}^{\mathbf{qq,kk}}_{\alpha \beta}  -\rho^{g_{\alpha}e}_{\mathbf{q}\mathbf{q}}  \rho^{eg_{\beta}}_{\mathbf{k}\mathbf{k}} \bar{\mathcal{G}}^{\mathbf{qq,kk}}_{\alpha \beta} \right) \\
   &+i \sum_k (\rho^{g_{1-\alpha} g_{\alpha}}_{\mathbf{q}\mathbf{q}}  \mathcal{G}_{01}^{\mathbf{k}\mathbf{q},\mathbf{q}\mathbf{k}} \rho^{ee}_{\mathbf{k}\mathbf{k}} - \rho^{g_{\alpha} g_{1-\alpha}}_{\mathbf{q}\mathbf{q}}  \bar{\mathcal{G}}_{01}^{\mathbf{k}\mathbf{q},\mathbf{q}\mathbf{k}} \rho^{ee}_{\mathbf{k}\mathbf{k}})\\
\end{split} 
\end{equation}

\begin{equation}
\begin{split}
\frac{d \rho^{g_{\alpha}e}_{\mathbf{q}\mathbf{q}}}{dt} &=  i\sum_{\beta \gamma} \sum_k (\rho^{g_{\alpha}g_{\gamma}}_{\mathbf{q}\mathbf{q}}-\delta_{\gamma \alpha}\rho^{ee}_{\mathbf{q}\mathbf{q}})  \mathcal{G}_{\gamma \beta}^{\mathbf{qq,kk}} \rho^{g_{\beta}e}_{\mathbf{k}\mathbf{k}} \\
  & + i \sum_{\beta} \sum_k \rho^{g_{\alpha}e}_{\mathbf{q}\mathbf{q}}  \mathcal{G}_{\beta \beta}^{\mathbf{k}\mathbf{q},\mathbf{q}\mathbf{k}} (\delta_{\alpha \beta} \rho^{ee}_{\mathbf{k}\mathbf{k}} - \rho^{g_{\beta}g_{\beta}}_{\mathbf{k}\mathbf{k}}) \\
  &+ i\sum_k\left[ (  -\rho^{g_{\alpha}e}_{\mathbf{q}\mathbf{q}}   \mathcal{G}_{01}^{\mathbf{k}\mathbf{q},\mathbf{q}\mathbf{k}} ( \rho^{g_{1-\alpha}g_{\alpha}}_{\mathbf{k}\mathbf{k}}  + \rho^{g_{\alpha}g_{1-\alpha}}_{\mathbf{k}\mathbf{k}} ) \right.\\
  &\quad \quad + \left. \rho^{g_{1-\alpha}e}_{\mathbf{q}\mathbf{q}}  \mathcal{G}_{01}^{\mathbf{k}\mathbf{q},\mathbf{q}\mathbf{k}} \rho^{ee}_{\mathbf{k}\mathbf{k}}) \right]\\
  &- \Gamma/2 \, \rho^{g_{\alpha}e}_{\mathbf{q}\mathbf{q}}
\end{split}
\end{equation}
and
\begin{equation}
\begin{split}
    \frac{d \rho^{g_{0}g_{1}}_{\mathbf{q}\mathbf{q}}}{dt} &=  - i \sum_k  \left(     \rho^{eg_1}_{\mathbf{q}\mathbf{q}} \mathcal{G}_{00}^{\mathbf{qq,kk}} \rho^{g_0e}_{\mathbf{k}\mathbf{k}}  - \rho^{g_0e}_{\mathbf{q}\mathbf{q}} \mathcal{\bar{G}}_{11}^{\mathbf{qq,kk}} \rho^{eg_1}_{\mathbf{k}\mathbf{k}} \right. \\
    &\qquad \qquad \left. +\rho^{eg_1}_{\mathbf{q}\mathbf{q}} \mathcal{G}^{\mathbf{qq,kk}}_{01} \rho^{g_1e}_{\mathbf{k}\mathbf{k}} -\rho^{g_0e}_{\mathbf{q}\mathbf{q}} \mathcal{G}^{\mathbf{qq,kk}}_{01} \rho^{eg_0}_{\mathbf{k}\mathbf{k}} \right) \\
     & \quad + i \sum_k \sum_{\beta}\rho^{g_0 g_1}_{\mathbf{q}\mathbf{q}} \mathcal{G}_{\beta \beta}^{\mathbf{q}\mathbf{k},\mathbf{k}\mathbf{q}} \rho^{ee}_{\mathbf{k}\mathbf{k}}\\
     & \quad + i\sum_k  (\rho^{g_1g_1}_{\mathbf{q}\mathbf{q}} - \rho^{g_0g_0}_{\mathbf{q}\mathbf{q}})\mathcal{G}_{01}^{\mathbf{q}\mathbf{k},\mathbf{k}\mathbf{q}} \rho^{ee}_{\mathbf{k}\mathbf{k}}
\end{split}
\label{eq_sup:rhogg}
\end{equation}
Here, $\mathcal{G}^{\mathbf{k}\mathbf{k},\mathbf{q}\mathbf{q}}_{\alpha \beta} =\Delta^{\mathbf{k}\mathbf{k},\mathbf{q}\mathbf{q}}_{\alpha \beta} +i \Gamma^{\mathbf{k}\mathbf{k},\mathbf{q}\mathbf{q}}_{\alpha \beta}$ . We further used that $\mathcal{G}_{\alpha\beta}^{\mathbf{k}\mathbf{q},\mathbf{k}^{\prime}\mathbf{q}^{\prime}} =\mathcal{G}_{\beta \alpha}^{\mathbf{k}\mathbf{q},\mathbf{k}^{\prime}\mathbf{q}^{\prime}} = \mathcal{G}_{\beta \alpha}^{\mathbf{k}^{\prime}\mathbf{q}^{\prime},\mathbf{k}\mathbf{q}} $ and $\sum_k \sum_{\alpha} \Gamma_{\alpha \alpha}^{\mathbf{q}\mathbf{k},\mathbf{k}\mathbf{q}} = \Gamma/2$ for the matrix elements in the geometry we consider with the explicit expressions derived next.

 \subsubsection{Evolution of the excited state population}
 Considering for a moment only the evolution of the excited state population in Eq.~\ref{eq_sup:rhoee}, we note that the first line corresponds to spontaneous decay of the excited state $e$ in momentum $\vect{q}$ to ground state $g_{\alpha}$ in momentum $\vect{k}$ mediated by $\Gamma^{\mathbf{k}\mathbf{q},\mathbf{q}\mathbf{k}}_{\alpha\alpha}$ and Pauli blocked by $1-\rho_{\mathbf{k}\mathbf{k}}^{g_{\alpha}g_{\alpha}}$. The second line in contrast can result in a sub/super-radiant decay mediated by the interactions $\mathcal{G}^{\mathbf{qq,kk}}_{\alpha \beta} $ and the coherences $\rho^{eg_{\alpha}}$. 
 
 For the initial states we consider we have $\rho^{g_0g_0} \sim \cos^2(\theta/2)$ which therefore controls the Pauli blocking factor. In contrast the relevant initial coherences are $\rho^{eg_0} \rho^{g_0e} \sim (\sin(\theta/2) \cos(\theta/2))^2$. Since we normalise the decay rate by the initial excited state population $\rho^{ee} \sim \sin^2(\theta/2)$, the interaction induced change scales as $\cos^2(\theta/2)$.

\subsection{Matrix elements for 2D homogeneous system}
In the strongly confined two dimensional limit an appropriate single-particle basis is given by $ \phi_i(x,y,z) = \frac{1}{\sqrt{A}} \psi_0(z) e^{i (q_{i,x} x + q_{i,y} y)}$, where $\psi_0(z) = \frac{1}{\sqrt{a_z \sqrt{\pi}}} e^{- (z/a_z)^2/2}$ is the ground state harmonic oscillator wave-function along $z$ and $a_z = \sqrt{\hbar/(M \omega_z)}$ the corresponding oscillator length, $e^{i (q_{i,x} x + q_{i,y} y)}$ are plane waves in the 2D plane, and $A$ is the two-dimensional area of the homogeneous system. In our simulations we use $\omega_z = 40$ kHz unless stated explicitly otherwise.

In this basis the matrix elements of the Green's function become [c.f.~Eqs.~(\ref{eq_sup:Delta_ijkl}) and (\ref{eq_sup:Gamma_ijkl})]
\begin{align}
    G^{\mathbf{ij,kl}} &= \frac{1}{A^2} \int d\vect{r}_{2D}  d\vect{r}_{2D}^{\prime}  e^{i \vect{r}_{2D} (\vect{q}_j-\vect{q}_i)}e^{i \vect{r}^{\prime}_{2D} (\vect{q}_k-\vect{q}_l)} \nonumber\\
    & \quad \int dz dz^{\prime}  |\psi_0(z)|^2  \, G(\vect{r}_{2D}- \vect{r}_{2D}^{\prime},z-z^{\prime}) \, |\psi_0(z^{\prime})|^2
\end{align}
Defining the second line as the two-dimensional Green's function $G^{2D}$ we then have
\begin{align}
G^{\mathbf{ij,kl}} &= \frac{1}{A^2} \int d\vect{r}_{2D}   e^{i \vect{r}_{2D} (\vect{q}_j-\vect{q}_i)} \nonumber\\
&\int d\vect{r}_{2D}^{\prime}  e^{i \vect{r}^{\prime}_{2D}(\vect{q}_l-\vect{q}_k)} G^{2D}(\vect{r}_{2D} - \vect{r}^{\prime}_{2D})\\
 &= \frac{1}{A} \delta_{\vect{q}_j -\vect{q}_i + (\vect{q_l}-\vect{q}_k)} \tilde{G}^{2D}(\vect{q}_j - \vect{q}_i) \label{eq_sup:M_matrix_elements}
\end{align}
with $\tilde{G}^{2D}(\vect{q}) = \int d\vect{r}_{2D}  e^{i \vect{r}_{2D} \vect{q}} \, G^{2D}(\vect{r}_{2D})$.

\subsubsection{Calculation of the 2D Green's function}
We can compute the 2D Green's function as
\begin{align}
    \tilde{G}^{2D}(\vect{q}) &= \int d\vect{r}_{2D}  e^{i \vect{r}_{2D} \vect{q}}  \, G^{2D}(\vect{r}_{2D}) \\
    &=\int d\vect{r}_{2D}  e^{i \vect{r}_{2D} \vect{q}} \nonumber\\
    &  \int dz dz^{\prime} |\psi_0(z)|^2  \, G(\vect{r}_{2D},z-z^{\prime}) \, |\psi_0(z^{\prime})|^2\\
    &=\int d\vect{r}_{2D}  e^{i \vect{r}_{2D} \vect{q}} \int dz dz^{\prime} \frac{dq_z}{2\pi} \frac{dq^{\prime}_z}{2\pi} e^{-i q_z z} e^{-i q_z^{\prime} z^{\prime}} \nonumber\\
    & \int  \mathcal{F}[|\psi_0|^2](q_z)  \, G(\vect{r}_{2D},z-z^{\prime}) \, \mathcal{F}[|\psi_0|^2](q^{\prime}_z)\\
    &= \int \frac{d q_z}{ 2 \pi} \mathcal{F}[|\psi_0|^2](q_z) \tilde{G}(q_x,q_y,q_z) \mathcal{F}[|\psi_0|^2](-q_z)
\end{align}
where we defined $\mathcal{F}[|\psi_0|^2] (q_z) = \int dz |\psi_0(z)|^2 e^{i q_z z}$ and $\tilde{G}(\vect{q}) = \int d^3\vect{q} \, e^{i \vect{q} \vect{r}} G(\vect{r})$.

\subsubsection{Fourier-transform of the Green's function}
We evaluate the Fourier transform of the 3D Green's function, $\tilde{G}_{R(I)}(\vect{q}) = \int d^3\vect{q} \, e^{i \vect{q} \vect{r}} G_{R(I)}(\vect{r})$, as
\begin{align}
    \tilde{G}_R(\vect{q}) &= \frac{(2 \pi)^3}{ 2 \pi^2}  \, \left[  \frac{2 k_0^2 + q^2}{3 k_0^3 (q^2-k_0^2)} (\mathbf{I} - 3 \vect{\hat{q}} \otimes \vect{\hat{q}}) \right. \nonumber\\
    & \quad \quad \quad \left. +  \frac{2}{k_0(q^2-k_0^2)} (\vect{\hat{q}} \otimes \vect{\hat{q}}) \right],
\end{align}
and
\begin{equation}
    \tilde{G}_I(\vect{q}) = \frac{(2 \pi)^3}{ 4 \pi} \frac{1}{k_0^2} \, (\mathbf{I} - \vect{\hat{q}} \otimes \vect{\hat{q}}) \, \delta(q-k_0). \label{sup_eq:GI_3D}
\end{equation}
These expressions allow the explicit evaluation of the two-dimensional Green's function $\tilde{G}^{2D}(\mathbf{q})$ as shown in the next subsection.

\subsubsection{Imaginary Part/Consequences of 2D on Pauli blocking}
Explicitly, the imaginary part of the 2D Green's tensor reads
\begin{equation}
    \tilde{G}^{2D}_I(\vect{q}) =   \frac{1}{2 \pi } \frac{1}{ k_0^3 \sqrt{k_0^2-q^2}}\left(k_0^2 \mathbf{I} - q^2 \vect{\hat{q}} \otimes \vect{\hat{q}}\right) \theta(k_0-q) \label{sup_eq:GI_2D}
\end{equation}
where $\mathbf{q}=(q_x,q_y,0)$ is the in-plane momentum and we dropped the factor $e^{-1/2 a_z^2 \left(k_0^2-q^2\right)}$ which reduces to 1 in the strongly confined limit $a_z k \rightarrow 0$ with the harmonic oscillator length $a_z = \sqrt{\hbar/(M \omega_z)}$

We point out the following important modification in the two-dimensional setting. Whereas in 3D [Eq.~(\ref{sup_eq:GI_3D})] the imaginary part couples states $\mathbf{q}$ to states $\mathbf{q}^{\prime}$ on a spherical shell with $|\mathbf{q} - \mathbf{q}^{\prime}| = k_0$, in 2D [Eq.~(\ref{sup_eq:GI_2D})] the coupling occurs to all states with $|\mathbf{q} - \mathbf{q}^{\prime}| \le k_0$.

To understand the relevance of this for Pauli blocking, consider a perfect Fermi sea at $T=0$ with $k_0> 2 k_f$. In 3D all states inside the Fermi sea then couple to states outside the Fermi sea, and no Pauli blocking can occur. However, in 2D for every state inside the Fermi sea at least some of the coupled states will always also be within the Fermi sea, resulting in some degree of Pauli blocking for arbitrary $k_0$.

\subsection{Approximating harmonically confined gas}
We have developed our master equation in a plane wave basis, which describes a homogenous system in a box potential. To facilitate comparisons of this model with a weakly trapped harmonically confined system, we have to properly choose the parameters of the box trap to adjust for the distinct densities of states.

Specifically, for a homogeneous 2D gas in a box of size $L \times L$, the density of states is constant with each momentum state occupying a volume of $((2 \pi)/L)^2$ in momentum space. The Fermi energy is $E_F = \hbar^2 k_F^2/(2M)$ in terms of the Fermi momentum $k_F$ and the atomic mass $M$. Given the constant density there are
\begin{equation}
    N = \frac{\pi k_F^2}{((2 \pi)/L)^2}
\end{equation}
particles within the Fermi sea.

In contrast, for a 2D harmonically confined gas the density of states is linear in energy $E$. The Fermi energy depends on the total atom number and trapping parameters as $E_F = \hbar \omega_{\perp} \sqrt{2N}$ for a radial trapping frequency $\omega_{\perp}$. For all simulations in this work we use $\omega_{\perp} = 150$ Hz.

To now make a direct comparison of the box system to the trapped system, we want to ensure that both have the same Fermi energy $E_F$ and the same number of particles $N$. Therefore, we choose the box size $L$ of the homogeneous system as
\begin{equation}
    L = \sqrt{ \frac{\pi k_{F}^2}{4 \pi^2 N}} = \sqrt{ \frac{ M E_F}{2 \pi \hbar^2 N}} \label{eq_sup:box_size}
\end{equation}
in terms of the Fermi energy $E_F$ of the harmonically trapped system, and the total atom number $N$.

At finite temperature the harmonically trapped gas expands further. To account for this effect we scale $L$ by an $N$ and $T$ dependent factor obtained from a semi-classical calculation. Using the semi-classical distribution function of a trapped Fermi gas, $f(r,p) = \frac{1}{\mathcal{Z}^{-1} \exp\left\{-\beta \left[\hbar^2 \omega_{\perp}^2r^2 + p^2/(2M)\right] \right\} +1}$, where $\mathcal{Z}$ is the fugacity ensuring normalisation $\int d^2 \vect{r} d^2 \vect{p} f_{\mathrm{FD}}(\vect{r},\vect{p}) = N$, we can compute the "average area" the Fermi gas occupies at finite temperature. Defining $\left< r^2 \right>(T,N) = \int d^2p \, d^2r \,r^2  f(r,p) /\int d^2p\, d^2r\, f(r,p) $ we rescale the box size $L$ by the thermal expansion factor $\sqrt{\left< r^2 \right>(T,N)/\left< r^2 \right>(0,N)}$.

\subsection{Exact HO matrix elements}
To estimate the accuracy of the plane wave approximation, we compare to calculations performed in the harmonic oscillator state basis. Thus, we need the matrix elements of the Green's function with respect to the 2D harmonic oscillator wave-functions $\phi_{\vect{n}}(X,Y)$, where $\vect{n}=(n_x,n_y)$ are the harmonic oscillator quantum numbers:
\begin{align}
   \Gamma^{\vect{n_1}\vect{n_2},\vect{n_3}\vect{n_4}}_{\alpha\beta} &=  \int d^2 \vect{r} d^2 \vect{r}^{\prime} \bar{\phi}_{\vect{n_1}}(X^{\prime},Y^{\prime}) \bar{\phi}_{\vect{n_2}}(X,Y) \notag\\
   &\qquad  \times \mathrm{Im}\, G_{\alpha\beta}^{2D}(r-r^{\prime}) \phi_{\vect{n_3}}(X,Y)\phi_{\vect{n_4}}(X^{\prime},Y^{\prime}) \nonumber\\
    &=\frac{1}{(2 \pi)^2}  \int d^2 \mathbf{q} \, \mathcal{F}\left[\phi_{\vect{n_1} }\phi_{\vect{n_4}} \right](\mathbf{q}) \, \mathrm{Im}\, G_{\alpha\beta}^{2D}(\mathbf{q})  \notag \\
    & \qquad \qquad \qquad \qquad  \mathcal{F}\left[\phi_{\vect{n_2}} \phi_{\vect{n_3}} \right](-\mathbf{q}) .
    \label{eq_sup:HO_exact}
\end{align}
These cannot be evaluated analytically in closed form, and are numerically extremely challenging to compute due to the fast oscillations in the harmonic oscillator wave-functions. However, we can reduce their computation to a semi-closed expression consisting of sums over analytically known terms similar to the calculation in \cite{KRb_squeezing}. We note that the required number of these elements grows as $n_{max}^4$, where $n_{max}$ is the largest included harmonic oscillator index, making simulations for a large number of atoms or finite temperature unfeasible.

\subsection{Semi-classics}

In Eq. 3 of the main text, the first line can be used to obtain the modification of the radiative decay due to Pauli blocking  for a non-interacting system assuming a plane wave basis in the transverse direction. Inspired by the semi-classical analysis described in Ref.~\cite{Shuve_2009} we can generalize our equation to describe the decay rate of an harmonically tapped gas. This yields the   following expression describing the decay rate of the excited state as
\begin{equation}
    \gamma_{\mathrm{eff}} = \frac{\sum_{\alpha}\int d^3 \vect{q}  \, d^3 \vect{k} \, d^3 \vect{r} \, f_{e}(\vect{k},\vect{r}) \Gamma_{\alpha \alpha}^{\mathbf{k}\mathbf{q},\mathbf{q}\mathbf{k}} \left[1-f_{g_{\alpha}}(\vect{q},\vect{r}) \right] }{ \int d^3 \vect{k} d^3 \vect{r} \, f_{e}(\vect{k},\vect{r}) },
    \label{eq_sup:SC_3D}
\end{equation}
where $\Gamma_{\alpha,\alpha}^{\mathbf{k}\mathbf{q},\mathbf{q}\mathbf{k}}$ describes spontaneous radiative decay from an atom in momentum $\hbar\vect{k}$ in the excited state to momentum $\hbar\vect{q}$ in ground state $\alpha$ while emitting a photon of momentum $\hbar(\vect{k-q})$. In 3D, we have that $\Gamma_{\alpha,\alpha}^{\mathbf{k}\mathbf{q},\mathbf{q}\mathbf{k}} \sim \delta(|\vect{k-q}|-k_0)$ [see Eq.~(\ref{sup_eq:GI_3D})], such that decay occurs into states $\vect{q}$ on a sphere of radius $k_0$ around the initial momentum $\vect{k}$. The distribution functions are directly related to the semi-classical distribution function for harmonically confined fermions in 3D
\begin{align}
    f_{\mathrm{FD}}&\,(\vect{q},\vect{r}) =\nonumber\\ &\,\frac{1}{\mathcal{Z}^{-1}\exp \left[-\beta\left(\sum_{i=x,y,z} M \omega_{i}^2 r_i^2/2 + \hbar^2 q_i^2/(2M)\right)\right]  +1}
\end{align}
where $\beta=1/(k_BT)$, $k_B$ is the Boltzmann constant, $M$ is the atomic mass, $\omega_i$ are the trapping frequencies and $\mathcal{Z}$ is the fugacity ensuring normalisation, $\int d ^3 \vect{q} d^3 \vect{r} f_{\mathrm{FD}}(\vect{q},\vect{r}) = N$.
These expressions reduce to the ones used in Ref.~\cite{Shuve_2009} for the corresponding initial states and level configurations considered there.

The initial Rabi excitation with a laser resonantly addressing the $g_0\to e$ transition with laser wave number $\vect{k}_L$ and pulse area $\theta$, leads to a shift of the initial distribution by a momentum $\vect{k}_L$ associated to the absorption of a laser photon.
The resulting distributions    can be written as  $f_{e}(\vect{q},\vect{r}) =  \sin(\theta)^2 f_{FD,0}(\vect{q}-\vect{k_L},\vect{r})$ and $f_{g_{0}} = \cos(\theta)^2 f_{FD,0}(\vect{q},\vect{r})$ and $f_{g_1} =f_{FD,1}(\vect{q},\vect{r})$, where the indices $0,1$ denote the different fugacities $\mathcal{Z}_0,\mathcal{Z}_1$ associated to the $g_0$ and $g_1$ initial populations in the case of imbalanced mixtures.

In 2D we instead have 
\begin{equation}
    \gamma_{\mathrm{eff}} = \frac{\sum_{\alpha}\int d^2 \vect{q}  \, d^2 \vect{k} \, d^2 \vect{r} \, f_{e}(\vect{k},\vect{r}) \Gamma_{\alpha \alpha}^{{\rm 2D}\, \mathbf{k}\mathbf{q},\mathbf{q}\mathbf{k}} \left[1-f_{g_{\alpha}}(\vect{q},\vect{r}) \right] }{ \int d^2 \vect{k} d^2 \vect{r} \, f_{e}(\vect{k},\vect{r}) }
     \label{eq_sup:SC_2D}
\end{equation}
where $\Gamma^{\rm 2D}$ now is the corresponding 2D expression and allows decay into all modes with $|\vect{q}-\vect{k}| \le k_0$  [see Eq.~(\ref{sup_eq:GI_2D})], and $f$ is the two-dimensional semi-classical distribution function. Since the laser is oriented perpendicular to the 2D plane, $\vect{k}_L \sim k_0 e_z$, there is no momentum transfer within the 2D plane and we have $f_{e}(\vect{q},\vect{r}) =  \sin(\theta)^2 f_{FD,0}(\vect{q},\vect{r})$, $f_{g_{0}} = \cos(\theta)^2 f_{FD,0}(\vect{q},\vect{r})$, and $f_{g_1} =f_{FD,1}(\vect{q},\vect{r})$.

\subsection{Extension to array of 2D pancakes}
It is straightforward to include a number of pancakes in our model. Every index $\mathbf{q}=(q_x,q_y)$ now also includes the $z$ lattice site, such that $\mathbf{q}_i=(q_x,q_y,n_i)$, and the single-particle basis becomes $\phi_{\mathbf{q}_i} = \frac{1}{\sqrt{A}}  e^{i(q_x x+q_y y)} \psi_0(z-n_i a) $ with the lattice spacing $a$.

We need to compute the generalised matrix-elements
\begin{equation}
\begin{split}
    G_{R/I}^{\mathbf{q}_i \mathbf{k}_j   \mathbf{Q}_l  \mathbf{K}_s} &= \int d^3\mathbf{r} d^3\mathbf{r}^{\prime}  \, \bar{\phi}_{\mathbf{q}_i}(\mathbf{r}) \phi_{\mathbf{k}_l}(\mathbf{r}) \, G_{R/I}(\mathbf{r} - \mathbf{r}^{\prime}) \, \\
    & \qquad \qquad \qquad \bar{\phi}_{\mathbf{Q}_l}(\mathbf{r}^{\prime}) \phi_{\mathbf{K}_s}(\mathbf{r}^{\prime})
    \end{split}
\end{equation}
for the cases $\mathbf{q}_i =\mathbf{k}_j  $, $ \mathbf{Q}_l = \mathbf{K}_s$ and $\mathbf{q}_i =\mathbf{Q}_l   $, $\mathbf{k}_j =\mathbf{K}_s$. In the latter case, the integrand contains the product $\psi_0(z-n_i a) \psi_0(z-n_j a) \psi_0(z^{\prime}-n_i a) \psi_0(z^{\prime}-n_j a) $ which is exponentially small for $i\neq j$ due to the Gaussian decay of the ground state oscillator wave-function, and can be  safely neglected for $a_z \ll a$, where $a_z$ is the spread of the ground state wave-function $\psi_0$. 

Therefore, the inclusion of the pancake degree of freedom only requires the additional terms with $\mathbf{q}_i =\mathbf{k}_j  $, $ \mathbf{Q}_l = \mathbf{K}_s$ . These reduce to
\begin{equation}
    G_{R/I}^{i i j j} = 1/A^2 \int d^3\mathbf{r} d^3\mathbf{r}^{\prime}  \psi^2_0(z) \, G_{R/I}(\mathbf{r} - \mathbf{r}^{\prime})  \psi^2_0(z^{\prime}-(n_i-n_j) a) ,
\end{equation}
which do not depend on the momentum degrees of freedom. Thus, these integrals  can simply be evaluated numerically for the discrete set of $n_i-n_j$.

\subsection{Model and density distributions in array of pancakes}
To model the experimental sequence, we start from the semi-classical density distribution $n_{\alpha}(x,y,z;N_{\alpha},\beta) = \int d^3\vect{p} f_{\mathrm{FD}} (\vect{r},\vect{p})$ of the 3D Fermi gas with the trapping frequencies given below at the inverse temperature $\beta$ with a total number of atoms $N_0,N_1$ in the ground states $g_0,g_1$. To determine the number of atoms $N_{\alpha,n_i}$ that end up in each pancake $n_i$ after switching on the optical lattice, we integrate the distribution to obtain $N_{\alpha,n_i} = \int_{(n_i-1/2) a}^{(n_i+1/2) a} dz \int dx dy \, n_{\alpha}(x,y,z;N_\alpha,\beta)$, where $a$ is the $z$ lattice spacing.

Having determined the atom number $N_{\alpha,n_i}$ in each pancake $n_i$, we then define the momentum plane wave basis in each pancake. The occupation of momentum modes is sampled from independent 2D Fermi-Dirac distributions in each individual pancake at the given temperature and atom number.

All results in the main text then refer to the total populations and decay rates summed over all pancakes.

\subsection{Experimental sequence}
Our experiment starts with the preparation of a degenerate Fermi gas as described in Ref. \cite{sonderhouse_thermodynamics_2020}. $\mathrm{^{87}Sr}$ atoms in all 10 nuclear spin states are evaporated to a temperature of 0.6 times the Fermi temperature with radial confinement of $\omega_r = 2 \pi \times 100$ Hz and axial confinement of $\omega_z = 2 \pi \times 500$ Hz. Optical pumping is used to remove atoms in $m_F = -9/2$ so that the number of atoms with spin $m_F = -9/2$ is a factor of 10 smaller than the number of atoms in $m_F = -7/2$. The atoms are then adiabatically loaded into a vertical 1D optical lattice. The radial trap frequency in the lattice is $\omega_r = 2 \pi \times 150$ Hz, and the axial trap frequency is $\nu_z = 44$ kHz which gives a Lamb-Dicke parameter along z of $\eta = \sqrt{\nu^p_{rec}/\nu_{z}} = 0.33$, where $\nu^p_{rec}$ is the recoil frequency of the probe laser. The vertical extent of the cloud is 4 $\mu$m so that roughly 10 pancakes are loaded with 1000 atoms in $m_F = -7/2$ in the central pancake at a Fermi energy of 320 nK. The measured optical depth of the $m_F=-9/2$ component is 0.8.

\begin{figure}[t!]
\centerline{\includegraphics[width=9.5cm]{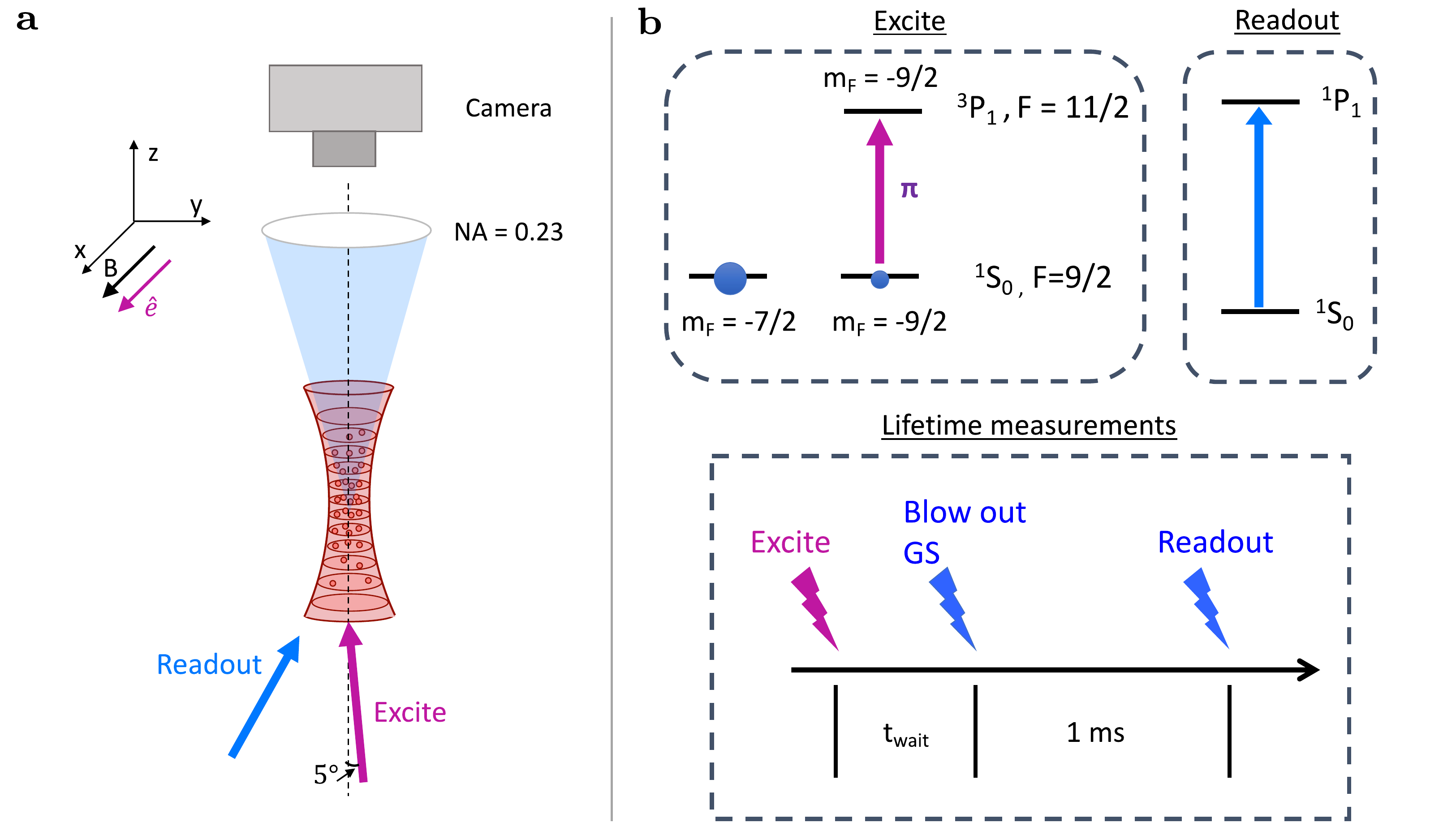}}
\caption{(a) Experimental setup. Light spin-selectively excites degenerate fermions trapped in a 1D optical lattice. The atomic population is then measured using a separate readout pulse, and the fluoresced light is detected using a high NA imaging system. (b) Readout scheme. Atoms in $\mathrm{^1S_0}$, $\mathrm{m_F = -9/2}$ are coherently excited to $\mathrm{^3P_1}$, $F=11/2$, $\mathrm{m_F} = -9/2$ using a short square pulse with pulse area $\pi$. After $t_{wait}$ the ground state population is removed and the number of excited atoms is read out after returning to the ground state via fluorescence imaging on the $\mathrm{^1S_0}$ - $\mathrm{^1P_1}$ transition.}
\label{fig:exp_fig}
\end{figure}

In order to directly measure spontaneous decay dynamics, we then coherently excite the $m_F = -9/2$ atoms on the $^1$S$_0$ - $\mathrm{^3P_1}$, $F=11/2$ transition at 689 nm. With its natural lifetime of 21.3 $\mu$s the $\mathrm{^3P_1}$ decay can be easily observed in a time-resolved fashion. An applied magnetic bias field of 3 G splits the excited state's sublevels so that the atoms can be selectively excited to the $m_F = -9/2$ state using a 5 $\mu$s $\pi$-pulse with $\pi$-polarized light. The probe beam is incident with 5 degrees with respect to z, as shown in Fig.~\ref{fig:exp_fig} (a).

To achieve fast, high signal-to-noise ratio measurements, the  atomic state readout is performed using the 30.4 MHz broad $\mathrm{^1P_1}$ transition, where roughly 100 photons can be scattered in 1 $\mu$s per atom. Around 1\% of the fluoresced photons are detected using a high resolution imaging axis with a numerical aperture  $\mathrm{NA} = 0.23$. The experimental sequence is shown in Fig.~\ref{fig:exp_fig} (b). First, the $m_F = -9/2$ atoms are excited to the $\mathrm{^3P_1}$, $F=11/2$, $\mathrm{m_F} = -9/2$ state as described above. After a variable time $t_{wait}$, a 10 $\mu$s pulse of high intensity $\mathrm{^1P_1}$ light causes significant recoil heating to the ground state atoms, which are as a result removed from the trap. After a further 1 ms, all excited atoms have decayed back to the ground state. A fluorescence image is then taken during a 2 $\mu$s long $\mathrm{^1P_1}$ readout pulse along the vertical camera axis. Wait times are varied from 1 $\mu$s to 200 $\mu$s, and 200 randomized measurements are recorded over 20 different wait times. The lifetime is then extracted from the time constant, which is determined by fitting the entire data set to a single exponential.

\end{appendix}
\end{document}